\documentclass[prd,tightenlines,floatfix,showpacs,preprintnumbers,nofootinbib,eqsecnum]{revtex4}

 \usepackage[dvips,final]{graphicx}
  \usepackage{amssymb}
   \usepackage{amsmath} 
    \usepackage{amsfonts}
     \usepackage{epsfig}
     \usepackage{color}
      \usepackage{bm} 
        \usepackage{xcolor}
        
\usepackage{booktabs}
\usepackage{slashed}

\def\doublespace{\def\baselinestretch{1.6}\large\normalsize}
\def\normalspace{\def\baselinestretch{1.0}\normalsize}
\def\PSfig#1#2{\scalebox{#1}{\includegraphics{#2}}}

\def\Caption#1{
  \normalspace
  \vskip-1mm\caption{\sl#1}\vskip-1mm
  \doublespace
}

\def\BA{\begin{eqnarray}}
\def\BE{\begin{equation}}
\def\BF{\begin{figure}[htb]}
\def\BT{\begin{table}[htb]}
\def\EA{\end{eqnarray}}
\def\EE{\end{equation}}
\def\EF{\end{figure}}
\def\ET{\end{table}}

\def\la{\langle}
\def\ra{\rangle}
\def\fm{\,\mbox{fm}}

\def\mb{\,\mbox{mb}}

\def\TeV{\,\mbox{TeV}}
\def\GeV{\,\mbox{GeV}}

\def\Jpsi{J\!/\!\psi}
\def\psip{\psi^{\,\prime}}

\def\Y{\Upsilon}
\def\Yp{\Upsilon^{\,\prime}}

\def\sqq{\sigma_{\bar QQ}}

\def\lsim{\mathrel{\rlap{\lower4pt\hbox{\hskip1pt$\sim$}}
     \raise1pt\hbox{$<$}}}         
 \def\gsim{\mathrel{\rlap{\lower4pt\hbox{\hskip1pt$\sim$}}
     \raise1pt\hbox{$>$}}}         

\begin{document}

\title{
Coherent photoproduction of heavy quarkonia on nuclei
}

\author{B. Z. Kopeliovich$^1$}
\email{boris.kopeliovich@usm.cl}

\author{M. Krelina$^{2,3}$}
\email{michal.krelina@cvut.cz}

\author{J. Nemchik$^{2,4}$}
\email{nemcik@saske.sk}

\author{I. K. Potashnikova$^1$}
\email{irina.potashnikova@usm.cl}


\affiliation{$^1$
Departamento de F\'{\i}sica,
Universidad T\'ecnica Federico Santa Mar\'{\i}a,
Avenida Espa\~na 1680, Valpara\'iso, Chile}
\affiliation{$^2$
Czech Technical University in Prague, FNSPE, B\v rehov\'a 7, 11519
Prague, Czech Republic}
\affiliation{$^3$
Physikalisches Institut, University of Heidelberg, Im Neuenheimer Feld 226, 69120 Heidelberg, Germany}
\affiliation{$^4$
Institute of Experimental Physics SAS, Watsonova 47, 04001 Ko\v sice, Slovakia
}

\date{\today}
\begin{abstract}
The differential cross section of coherent photo-production of heavy quarkonia on nuclear targets is calculated within the QCD color dipole formalism. 
The higher-twist nuclear shadowing corresponding to the $\bar QQ$ Fock component of the photon, is calculated including the correlation between dipole orientation $\vec r$ and impact parameter of a collision $\vec b$, which is related to the transverse momentum transfer  via Fourier transform. 
We also included  the leading twist gluon shadowing corresponding to higher Fock components of the photon  containing gluons, which have specifically short coherence time, especially for multi-gluon components, even at very high energies. The contribution of such fluctuating gluonic dipole is calculated employing the path-integral technique.
Our results  are in good agreement with recent ALICE data on charmonium production in ultra-peripheral nuclear collisions.
\end{abstract}


\maketitle

%
%
%
\section{Introduction}
\label{intro}
%
%
%

Exclusive photo-production of vector mesons on protons and nuclei has been an important playground for strong interaction models since early years of vector dominance model \cite{Stodolsky:1966am,Ross:1965qa} based on the pole dominated dispersion relation for the production amplitude (see the comprehensive review \cite{Bauer:1977iq}).
While for light mesons that assumption was reasonably justified by closeness of the meson pole to the physical region, photo-production  of heavy quarkonia, 
$J/\psi$, $\Upsilon$, etc. (or high $Q^2$ electro-production), cannot be dominated by the pole, which is far away from physical region, and other singularities, poles and cuts, become essential \cite{Hufner:1995qe,Hufner:1997jg}. Besides, photo-production on nuclei provide unique information about the space-time pattern of interaction \cite{Hufner:1996dr}.

With advent of the quantum-chromodynamics (QCD) era, an alternative description in terms of color dipoles was proposed in \cite{Kopeliovich:1981pz}, and applied to exclusive photo-production of heavy quarkonia on protons and nuclei \cite{Kopeliovich:1991pu,Hufner:2000jb,Ivanov:2002kc}.  

Experiments with ultra-peripheral collisions (UPC) at the Relativistic Heavy Ion Collider (RHIC) and Large Hadron Collider (LHC) (see e.g. \cite{Bertulani:2005ru}) offer new opportunities for measurements of exclusive photo-production of vector
mesons on protons and nuclei. Corresponding calculations, performed within the color-dipole approach \cite{Ivanov:2007ms,Kopeliovich:2020has}, as well as other theoretical descriptions of coherent production of heavy quarkonia in UPC, require  improvements. 
In the present paper, we aim  to minimize  theoretical uncertainties of the QCD dipole formalism for coherent production on nuclear targets. 

The  effect, missed in many calculations, is the correlation between the dipole transverse orientation $\vec r$ and the impact parameter of the collision $\vec b$. 
Since a colorless $\bar QQ$ dipole interacts only due to the difference between the $Q$ and $\bar Q$ scattering amplitudes, the dipole-proton interaction vanishes when dipole separation vector $\vec r$ is perpendicular to $\vec b$, while reaches maximal strength when they are parallel \cite{Kopeliovich:2007fv}. The influence of the $\vec r$-$\vec b$ correlation on the  differential cross section of exclusive photo-production of heavy quarkonia on protons was studied recently in \cite{Kopeliovich:2021dgx}.  
A significant difference was found with the description based on the conventional $b$-dependent dipole models, lacking the $\vec b$-$\vec r$ correlation,
like  b-IPsat \cite{Rezaeian:2012ji}, b-Sat \cite{Kowalski:2003hm}, b-CGC \cite{Kowalski:2006hc,Rezaeian:2013tka} and b-BK \cite{Cepila:2018faq} models,
for example.
We found that the correlation is especially important for photo-production of radially excited charmonium states due to the nodal structure of the wave function. This provides a stringent test of various models for $b$-dependent dipole amplitude. 
The importance of color dipole orientation in other processes has been discussed in Ref.~\cite{Kopeliovich:2021dgx} (see references therein).

However, on nuclear targets the effect of $\vec b$-$\vec r$ correlation is diluted, except the nuclear periphery, where the nuclear density steeply varies with $\vec b$. Therefore the impact of $\vec b$-$\vec r$ correlation on the azimuthal asymmetry of photons and pions turns out to be rather small \cite{Kopeliovich:2008dy,Kopeliovich:2007sd,Kopeliovich:2007fv,Kopeliovich:2008nx}.  Nevertheless, in the present paper, we implement the dipole orientation effect to minimize the theoretical uncertainties. The calculations rely on the realistic model for the $\vec r$ and $\vec b$ dependent partial $\bar QQ$-proton and $\bar QQ$-nucleus amplitudes proposed in \cite{Kopeliovich:2021dgx}, 
with parameters corresponding to GBW \cite{GolecBiernat:1998js,GolecBiernat:1999qd} and BGBK \cite{Bartels:2002cj} saturation models for the dipole cross section.

The present paper is organized as follows. 
The basics of the dipole formalism for photoproduction on a proton target, $\gamma^*p\to Vp$, is described in Sec.~\ref{dipole}. In particular, the effect of $\vec b$-$\vec r$ correlation, including explicit expression for the partial dipole amplitude is presented
in Sec.~\ref{r-b}.

The important part of the calculations, the $V\to \bar QQ$ distribution function in the light-front (LF) frame, is discussed in Sec.~\ref{dist-fun}, where the procedure of Lorentz boost from the rest frame of the quarkonium is described.  It is confronted with the unjustified, but popular photon-like structure for the $V\to \bar QQ$ transition. Its Lorentz-invariant form leads to a huge weight of the $D$-wave component in the quarkonium rest frame, contradicting the solutions of the Schr\"odinger equation with realistic potentials. 

The dipole formalism for coherent photoproduction on nuclei is formulated in Sec.~\ref{sec-formulas}.

Since the higher-twist nuclear effects for photoproduction of heavy quarkonia are very small, the main nuclear effects are related to gluon shadowing, considered in Sec. \ref{sec-gs}.
In terms of the Fock state expansion, that is related to the higher Fock components of the projectile photon, which contain besides the $\bar QQ$, additional gluons. The lifetime (coherence time) of such components turn out to be rather short, even for the very first state $|\bar QQg\ra$, which we calculate in Sec.~\ref{1-g} relying on the path-integral
formalism. In Sec.~\ref{n-g} we evaluate the coherence length for multi-gluon Fock states and demonstrate that it is dramatically reduced with addition of every extra gluon. We conclude that the single-gluon approximation is rather accurate in the available energy range.

Eventually, in Sec.~\ref{sec-data} we present  predictions for differential
cross sections $d\sigma/dt$ of coherent charmonium photoproduction, which are in a good accord with available results of the ALICE experiment, extracted from data on UPC. Here we present also predictions for other quarkonium states, like $\psip, \Y$ and $\Yp$. 

%
%
%
\section{Dipole formalism for photo-production of heavy quarkonia}
\label{dipole}
%
%
%

The color dipole formalism leads to the following factorized form of the dipole-nucleon photo-production amplitude \cite{Kopeliovich:1991pu},  
%
\BE
\mathcal{A}^{\gamma^{\ast} p\to V p}(x,\vec q\,)
=
\bigl\la V |\tilde{\mathcal{A}} |\gamma^*\bigr\ra
= 2\,\int d^2b\,e^{i\vec q\cdot\vec b}
\int d^2r\int_0^1 d\alpha\,
\Psi_{V}^{*}(\vec r,\alpha)\,
\mathcal{A}^N_{\bar QQ}(\vec r, x, \alpha,\vec b\,)\,
\Psi_{\gamma^\ast}(\vec r,\alpha,Q^2)\,.
\label{amp-p0}
\EE
%
Here $\vec q$ is transverse component of momentum transfer; $\alpha$ is the fractional light-front momentum carried by a heavy quark or antiquark of the $\bar QQ$ Fock component of the photon, with the transverse separation $\vec r$. That is the lowest Fock state, while the higher Fock components contribute by default to the dipole-proton amplitude. However, on a nuclear target these higher Fock components will be taken into account separately, due to coherence effect in gluon radiation. 

The dipole-proton amplitude  $\mathcal{A}^N_{\bar QQ}(\vec r, x, \alpha,\vec b\,)$ in Eq.~(\ref{amp-p0}) depends on the transverse dipole size $\vec r$ and impact parameter of collision $\vec b$. The essential feature of this amplitude is the $\vec r$-$\vec b$ correlation, explicitly presented below
in Sec.~\ref{r-b}. The LF distribution functions $\Psi_{\gamma^\ast}(r,\alpha,Q^2)$ and $\Psi_V(r,\alpha)$ correspond to the transitions $\gamma^\ast\to \bar QQ$ and $\bar QQ\to V$, respectively. They are specified below in Sect.~\ref{dist-fun}.

The amplitude (\ref{amp-p0}) depends on Bjorken $x$ evaluated in \cite{ryskin} in the leading $\log(1/x)$ approximation. Each radiated gluon provides a factor $\int_x^1 dx^\prime/x^\prime = \ln(1/x)$, where $x$ is the minimal value of the fractional LF gluon momentum related to the longitudinal momentum transfer $q_L=(M_V^2+Q^2)/2E_{\gamma}$. So,
\BE
x=\frac{M_V^2+Q^2}{s}=
\frac{M_{V}^2+Q^2}{W^2+Q^2-m_N^2}\,.
\label{x}
\EE

As far as the production amplitude (\ref{amp-p0}) is known, we are in a position to calculate the differential cross section,
\BE
\frac{d\sigma^{\gamma^\ast p\to V p}}{dq^2}
=
\frac{1}{16\,\pi}\,
\Bigl |
\mathcal{A}^{\gamma^\ast p\to V p}(x,\vec q\,)
\Bigr |^2\,.
\label{proton}
\EE

\subsection{Dipole-proton partial amplitude with 
\boldmath$\vec r$-$\vec b$ correlation}
\label{r-b}

As was explained in Introduction, interaction of a color neutral dipole with impact parameter $\vec b$ exhibits a strong $\vec{r}$-$\vec{b}$ correlation. Namely, interaction  vanishes if $\vec{r}\perp\vec{b}$, but reaches maximal strength if $\vec{r}\parallel\vec{b}$. An explicit form of the dipole-proton partial amplitude, possessing  such a correlation, was proposed in
 \cite{Kopeliovich:2008dy,Kopeliovich:2007sd,Kopeliovich:2007fv,Kopeliovich:2008nx,Kopeliovich:2021dgx} and reads,
%
\BA
\mathrm{Im} \mathcal{A}^N_{\bar QQ}(\vec r, x, \alpha,\vec b\,)
=
\frac{\sigma_0}{8\pi \mathcal{B}(x)}\,
\Biggl\{
\exp\left[-\,\frac{\bigl [\vec b+\vec
r(1-\alpha)\bigr ]^2}{2\mathcal{B}(x)}\right] 
+ 
\exp\left[-\,\frac{(\vec
b-\vec r\alpha)^2}{2\mathcal{B}(x)}\right]
\nonumber\\
- \,2\,\exp\Biggl[-\,\frac{r^2}{R_0^2(x)}
-\,\frac{\bigl [\,\vec b+(1/2-\alpha)\vec
r\,\bigr ]^2}{2\mathcal{B}(x)}\Biggr]
\Biggr\}\,,
\label{dipa-gbw}
 \EA
%
where the function $\mathcal{B}(x)$ was defined in Ref.~\cite{Kopeliovich:2008nx},
\BE
 \mathcal{B}(x)=B^{\bar qq}_{el}(x, r\to 0)-
 \frac{1}{8} R_0^2(x)\,.
 \label{slope}
 \EE
Here $R_0^2(x)$ controls $x$ dependence of the saturation cross section, introduced in \cite{GolecBiernat:1998js,GolecBiernat:1999qd}, $\sqq(r,x)=\sigma_0\,\bigl (1 - \exp \bigl [ - {r^2}/{R_0^2(x)}\bigr ] \bigr )$, with $\sigma_0 = 23.03\,\mb$, $R_0(x) = 0.4\,\fm\times(x/x_0)^{0.144}$ with $x_0 = 3.04\times 10^{-4}$. The dipole-proton slope in the limit of vanishingly small dipoles $B^{\bar qq}_{el}(x, r\to 0)$ can be measured in electro-production of vector mesons with highly virtual photons $Q^2\gg 1\GeV^2$. The measured slope $B_{\gamma^*p\to\rho p}(x,Q^2\gg 1\GeV^2)\approx 5\GeV^{-2}$ \cite{ZEUS:2007iet}, as expected is defined by the proton charge radius.

The GBW model, lacking DGLAP evolution, was improved in the BGBK dipole model \cite{Bartels:2002cj}, where the saturation scale is related to the gluon density, which is subject to DGLAP evolution, $R_0^2(x,\mu^2) = {4}/{Q_s^2(x,\mu^2)}={\sigma_0\,N_c} / \bigl ({\pi^2\,\alpha_s(\mu^2)\,x\,g(x,\mu^2)}\bigr )$, where $\mu^2 = {\mathrel{C}} / {r^2} + \mu_0^2$, $Q_s^2$ is the saturation scale and the gluon distribution function $x\,g(x,\mu^2)$ is obtained as a solution of the DGLAP evolution equation with the shape at the initial scale $Q_0^2 = 1\,\GeV^2$, $x\,g(x,Q_0^2)=A_g \,x^{-\lambda_g} (1-x)^{5.6}$. Here $A_g = 1.2$, $\lambda_g = 0.28$, $\mu_0^2 = 0.52\,\GeV^2$, $\mathrel{C} = 0.26$ and $\sigma_0 = 23\,\mb$. In what follows, we denote GBW and BGBK dipole model containing the color dipole orientation as br-GBW and br-BGBK, respectively.

\subsection{The \boldmath$\bar QQ$ distribution functions}
\label{dist-fun}

Now we should specify the form of the $\gamma^*\to\bar QQ$ and $V\to\bar QQ$ distribution functions in Eq.~(\ref{amp-p0}) resp. (\ref{amp-A0}). The former is well known \cite{Kogut:1969xa,Bjorken:1970ah}, here we skip the details, which can be found e.g. in \cite{Hufner:2000jb}).

The LF distribution function for heavy quarkonium $V\to\bar QQ$ requires special care. It is known in the rest frame of the quarkonium  treating it as a non-relativistic system and solving Schr\"odinger equation. However, challenging is the Lorentz boost to the LF frame.
A popular prescription was proposed by Terent'ev \cite{Terentev:1976jk}. The 3-dimensional quarkonium wave function in the rest frame can be expressed in terms of LF variables, fractional LF momentum $\alpha$ and 2-dimensional transverse momentum of the quarks.
Although the variables are invariant relative Lorentz boost, the function itself is not.
The crucial but {\it unjustified} assumption of the prescription is Lorentz invariance of the whole function of $\alpha$ and $p_T$, so it is boosted unchanged from the rest to the LF frame. 

The Lorentz boosted Schr\"odinger equation was derived in \cite{Kopeliovich:2015qna} for heavy $\bar QQ$ system, basing on the Lorentz-invariant Bethe-Salpeter equation and smallness of $\Delta\alpha=\alpha-1/2$. Comparison with the Terent'ev prescription (see fig.~2 in \cite{Kopeliovich:2015qna}) shows that the latter is amazingly accurate at medium values of $\alpha\sim 1/2$. Thus, the Lorentz boosting Terent'ev prescription is proven to be well justified. 

The quarkonium wave 
function in the rest frame was found solving the Schr\"odinger equation with several realistic  potentials, such as the harmonic oscillatory potential (HAR) (see e.g. \cite{Kopeliovich:1991pu,Cepila:2019skb}), Cornell potential (COR) \cite{Eichten:1978tg,Eichten:1979ms}, logarithmic potential (LOG) \cite{Quigg:1977dd}, power-like potential (POW) \cite{Barik:1980ai}, as well as Buchm\"uller-Tye potential (BT) \cite{Buchmuller:1980su}. All of them well reproduce the charmonium mass spectrum and leptonic decay widths. The results for the photoproduction cross sections obtained with these potentials were found in \cite{Hufner:2000jb} to be rather close. The observed diversity can be treated as a measure of theoretical uncertainty related to the LF quarkonium distribution function.

Nevertheless, a significant correction to this boosting prescription was found in \cite{Hufner:2000jb}. Because of transverse motion of the quarks, their momenta are not parallel to the boost axis, what leads to a quark spin rotation along with the boost, named after Melosh \cite{Melosh:1974cu} (see also \cite{Krelina:2018hmt,Lappi:2020ufv}). The corresponding correction to the photo-production cross section turns out to be large \cite{Hufner:2000jb} especially for $\psi(2S)$, what solves the $\psi^\prime$ to $J/\psi$ puzzle \cite{Hoyer:1999xe}.

On the contrary, the frequently used alternative assumption is that the $V\to\bar QQ$ vertex has the vector current structure $\Psi_{\mu}\bar u \gamma_{\mu}u$, like for $\gamma^*\to\bar QQ$ (see e.g. \cite{Ryskin:1992ui,Brodsky:1994kf,Frankfurt:1995jw,Nemchik:1996cw}). This vertex is Lorentz
invariant and can be treated in any reference frame, in particular in the $\bar QQ$ rest frame, where the three-dimensional Schr\"odinger equation discriminates between the state with different orbital momenta, in particular between the $S$- and $D$-waves \cite{Hufner:2000jb}. The above photon-like vertex contains a $D$-wave component with abnormally large 
weight, 5-6\% for charmonium and 1.6-1.8\% for bottomonium (uncertainty is related to the choice of renormalization-scheme dependent quark masses). In contrast, the solution of Schr\"odinger equation with above mentioned realistic potentials, contains an order of magnitude smaller contamination of the $D$-wave component 
(for a review on quarkonium physics see e.g. \cite{Brambilla:2004wf,Brambilla:2010cs}).

The calculations in \cite{Haysak:2003yf}, employing the COR potential and hyperfine splitting, exhibit $D$-wave contributions to the $J/\psi$ wave function squared with magnitudes $0.05\%$, 
and $0.004\%$ for $\Upsilon(1S)$.

Similar magnitude of the $D$-wave component for charmonium was found in Ref.~\cite{Fu:2018yxq}, and in Ref.~\cite{Chang:2010kj} in the framework of the Bethe-Salpeter equation, as well as in Ref.~\cite{Cao:2012du} within the QCD-inspired quark potential model.

These solid arguments completely rule out the possibility of photon-like vertex for transition of a heavy quarkonium to a $\bar QQ$ pair. For this reason we treat this unjustified model as incorrect and ignore it in what follows.

%
%
%
\section{Coherent  photo-production off nuclei: higher twist shadowing}
\label{sec-formulas}
%
%
%

The lowest Fock component of the projectile photon, which contributes to this process, is $\bar QQ$. The transverse dipole size is small, $1/m_Q$. Therefore shadowing corrections are as small as $1/m_Q^2$, so should be treated  as a higher twist effect.

For the amplitude of quarkonium photo-production on a nuclear target, $\gamma^*A\to VA$, one can employ expression Eq.~(\ref{amp-p0}), but replacing the dipole-nucleon by dipole-nucleus amplitude,
\BE
\mathcal{A}^{\gamma^{\ast} A\to V A}(x,Q^2,\vec q\,)
=
2\,
\int d^2b_A\,e^{i\vec q\cdot\vec b_A}
\int d^2r\int_0^1 d\alpha\,
\Psi_{V}^{*}(\vec r,\alpha)\,
\mathcal{A}^A_{\bar QQ}(\vec r, x, \alpha,\vec b_A)\,
\Psi_{\gamma^\ast}(\vec r,\alpha,Q^2)\,.
\label{amp-A0}
\EE

In ultra-peripheral collisions at the LHC, the photon virtuality $Q^2\sim 0$ and the photon energy in the nuclear rest frame is sufficiently high to make the coherence length $l_c$ for $\bar QQ$ photo-production much longer than the nuclear radius, $l_c = 1/q_L=(W^2+Q^2-m_N^2)/\bigl (m_N\,(M_V^2+Q^2)\approx W^2/(m_N M_V^2)\bigr)\gg R_A$.
Then Lorentz time delation freezes the fluctuations of the dipole size, and one can rely on
the eikonal form for the dipole-nucleus partial amplitude at impact parameter $\vec {b}_A$, 
\BE
\mathrm{Im} \mathcal{A}^A_{\bar QQ}(\vec r, x, \alpha,\vec b_A)\Biggl |_{l_c\gg R_A}
=
1 - \Biggl [1 - \frac{1}{A}\,
\int d^{2} b\,\,
\mathrm{Im} \mathcal{A}^N_{\bar QQ}(\vec r, x, \alpha, \vec{b}\,)\,
T_{A}(\vec{b}_A+\vec{b}\,)
\Biggr ]^A\,,
\label{eik}
\EE
where $T_A(\vec b_A) = \int_{-\infty}^{\infty} dz\,\rho_A(\vec b_A,z)$ is the nuclear thickness function normalized as $\int d^2 b_A\,T_A(\vec b_A) = A$, and $\rho_A(\vec b_A,z)$ is the nuclear density. For $\rho_A(\vec b_A,z)$ we employ the realistic Wood-Saxon form \cite{DeJager:1987qc}.

Expression for the differential cross sections is analogous to Eq.~(\ref{proton}) provided that the coherence length $l_c\gg R_A$,
%
\BE
\frac{d\sigma^{\gamma^\ast A\to V A}(x,Q^2,t=-q^2\,)}{dt}\Biggl |_{l_c\gg R_A}
=
\frac{1}{16\,\pi}\,
\Bigl |
\mathcal{A}^{\gamma^\ast A\to V A}(x,Q^2, \vec q\,)
\Bigr |^2\,.
\label{nucleus}
\EE

We also incorporate the real part \cite{Bronzan:1974jh,Nemchik:1996cw,forshaw-03} of the nuclear $\gamma^\ast A\to V A$ amplitude via a substitution in 
Eq.~(\ref{eik})
\BE
\mathrm{Im}
\mathcal{A}_{\bar QQ}^{N}(\vec r, x, \alpha, \vec{b}\,)
\Rightarrow
\mathrm{Im}
\mathcal{A}_{\bar QQ}^{N}(\vec r, x, \alpha, \vec{b}\,)
\,\cdot
\left(1 - i\,\frac{\pi\,\Lambda}{2}\right)\,
\label{bronzan}
\EE
with 
$\Lambda=\partial\ln(\,
{\mathrm{Im}\mathcal{A}^{N}_{\bar QQ}(\vec r,x,\alpha,\vec b\,)}\,)
\huge/\partial\ln (1/x)$.

In order to include the skewness correction \cite{Shuvaev:1999ce} one should perform the following modification of the partial $\bar QQ$-nucleon amplitude in the eikonal formula (\ref{eik}),
%
\BA
\mathrm{Im} \mathcal{A}^N_{\bar QQ}(\vec r, x, \alpha,\vec{b}\,)\Rightarrow
\mathrm{Im} \mathcal{A}^N_{\bar QQ}(\vec r, x, \alpha,\vec{b}\,)\cdot 
R_S(\Lambda)\,,
\label{skewness-a}
\EA
%
where the skewness factor 
$R_S(\Lambda)=\bigl ({2^{2\Lambda + 3}}/{\sqrt{\pi}}\, \bigr) \,\cdot {\Gamma(\Lambda + 5/2)}/{\Gamma(\Lambda + 4)}$.

%
%
%
\section{Higher Fock states and gluon shadowing}
\label{sec-gs}
%
%
%

Higher Fock components of the photon contain besides the $\bar QQ$ pair additional gluons,
$|\bar QQ\,g\ra$,  $|\bar QQ\,2g\ra$, etc. In $\gamma^*p$ collision they correspond to gluon radiation processes, which should be treated as higher-order corrections to the gluonic exchange, which take part in the building of the Pomeron exchange in the diffractive $\gamma^*p$ interaction. Therefore the higher Fock components are included in the $\bar QQ$-dipole interaction with the proton. In the case of photo-production on a nucleus, the higher Fock components contribute to the amplitude $\mathcal{A}^N_{\bar QQ}(\vec r, x, \alpha, \vec{b})$ in Eq.~(\ref{eik}). This would correspond to the Bethe-Heitler regime of radiation, when each of multiple interactions produce independent gluon radiation without interferences.

However, the pattern of multiple interactions changes in the regime of long coherence length of gluon radiation $l_c^g\gg d$, where $d\approx 2\fm$ is the mean separation between bound nucleons. The gluon radiation length has the form,
\BE
l_c^g=\frac{2E_{\gamma}\alpha_g(1-\alpha_g)}{k_T^2+(1-\alpha_g)m_g^2+\alpha_g M_{\bar QQ}^2}, 
\label{lg-full}
\EE
where $\alpha_g$ is the LF fraction of the photon momentum carried by the gluon,
$M_{\bar QQ}$ is the effective mass of the ${\bar QQ}$ pair and 
the effective gluon mass  $m_g\approx0.7\GeV$ is fixed by data on gluon radiation \cite{Kopeliovich:1999am,Kopeliovich:2007pq}. Such a rather large effective gluon mass makes the gluonic coherence length (see Eq.~(\ref{lg}) in Sec.~\ref{n-g}) nearly order of magnitude shorter than for $\bar QQ$ fluctuations \cite{Kopeliovich:2000ra}. This is why onset of shadowing in DIS, which occur at $x<0.1$ for higher twist $\bar QQ$ shadowing, requires much smaller $x<0.01$ for onset of gluon shadowing \cite{Kopeliovich:1999am}. That was confirmed by the global analysis of DIS data on nuclei \cite{florian}.

\subsection{Single gluon approximation}
\label{1-g}

At long $l_c^g\gg d$ the Landau-Pomeranchuk effect is at work, namely, radiation does not resolve multiple interactions, which act as one accumulated kick, which shakes off gluons. So  gluon radiation intensity is reduced in comparison with the Bethe-Heitler regime. This is why it is called {\it gluon shadowing} (GS) \cite{Kopeliovich:1999am}.

Such a reduction of the cross section is a well known feature of Gribov inelastic shadowing \cite{Gribov:1968jf,Kopeliovich:2012kw,Kopeliovich:2016jjx}, and gluon shadowing is a part of Gribov corrections. It correspond to inclusion of higher Fock components of the projection photon, $|\bar QQg\ra$, $|\bar QQ2g\ra$, etc. Then, eikonalization of these components is required. However, differently from $\bar QQ$ fluctuations, a $\bar QQg$ component does not reach the "frozen" size regime even at very small $x$, because of divergent $d\alpha_g/\alpha_g$ behavior. Therefore, variation of the $\bar QQ-g$ dipole size should be taken into account. 
This was done for DIS in \cite{Kopeliovich:1999am} applying the Green function technique. It was also used for calculation of gluon shadowing in electroproduction of heavy vector mesons on nuclei in  Refs.~\cite{Ivanov:2002kc,Nemchik:2002ug}. 

The Gribov correction, related to the $\bar QQg$ component of the photon, to the partial nuclear cross section at impact parameter $b_A$, reads
\BA
\Delta \sigma^{\gamma^*A}_{tot}(b_A)
&=&
\frac{1}{2} \,{\rm Re}
\int\limits_{-\infty}^{\infty} dz_2\,\rho_A(b_A,z_2)
\int\limits_{-\infty}^{z_2} dz_1\,\rho_A(b_A,z_1)
\int d^2\rho_1\int d^2\rho_2
\int \frac{d\alpha_g}{\alpha_g}\nonumber\\
&\times&
 A^\dagger_{\gamma^*\to \bar QQg}(\alpha_g,\vec\rho_2)
G_{gg}(z_2,\vec\rho_2;z_1,\vec\rho_1)
A_{\gamma^*\to \bar QQg}(\alpha_g,\vec\rho_1)\,.
\label{Delta-sig}
\EA
This is the usual Gribov correction containing product of conjugated  amplitudes of diffractive transitions, $\gamma^*+N\to \bar QQg+N$, on bound nucleons with longitudinal coordinates $z_{1,2}$. On the contrary to \cite{kk} we have no uncertainty with inclusion of higher order multiple interactions. The propagation of the $\bar QQg$ system in the nuclear medium is described by the Green function $G_{gg}(z_2,\vec\rho_2;z_1,\vec\rho_1)$, which takes care of attenuation and phase shift.

The $\bar QQg$ Fock state is characterized by two scales. One scale is set by the heavy quark mass, and determines the small $\bar QQ$ separation $\approx 1/m_Q$. This is a higher twist effect, i.e. the quark $\bar QQ$ pair size drops linearly with $m_Q$ (or with $Q^2$ in \cite{Kopeliovich:1999am}), so it can be treated as point-like color-octet system. However, the transverse distance between the $\bar QQ$ and gluon is much larger and independent of $m_Q$ (up to Log corrections). This distance depends on the previously mentioned scale, the effective gluon mass, $m_g\approx 0.7\GeV$. Thus, the $\bar QQ-g$ system is strongly asymmetric, a nearly point-like color-octet $\bar QQ$ pair and the gluon is separated by relatively long, $1/m_g$ distance. Since the latter controls gluon shadowing, it is the leading twist effect, because it is hardly dependent (only Log) on the hard scale $m_Q$. Expression (\ref{Delta-sig}) is integrated over $\bar QQ$ separation, assuming a weak dependence of the Green function on this scale. Thus, the whole 3-body system can be treated with high precision as glue-glue dipole  \cite{Kopeliovich:1999am}. We skip further details of calculation, which are well described in \cite{Kopeliovich:1999am,Ivanov:2002kc}.

The fractional Gribov correction to the partial nuclear cross section
\BE
R_G(b_A)=1-\frac{\Delta \sigma^{\gamma^*A}_{tot}(b_A)}{T_A(b_A)
\sigma^{\gamma^*N}_{tot}},
\label{RG}
\EE
is interpreted in the parton model as ratio of gluon densities. As far as the gluon density in the bound nucleons is reduced by factor $R_G$ one can take this into account renormalizing the nucleon amplitude in Eq.~(\ref{eik}),
\BE
\mathrm{Im} \mathcal{A}^N_{\bar QQ}(\vec r, x, \alpha,\vec b)
\Rightarrow 
\mathrm{Im} \mathcal{A}^N_{\bar QQ}(\vec r, x, \alpha,\vec b)
\cdot R_G(x,|\vec b_A +\vec b|)\,.
\label{eq:dipole:gs:replace-b}
\EE

\begin{figure}[hbt]
    \includegraphics[height=6cm]{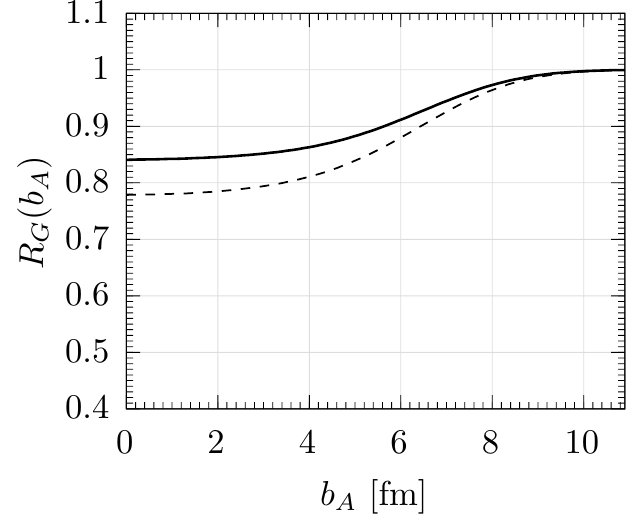}~~~~
    \includegraphics[height=6cm]{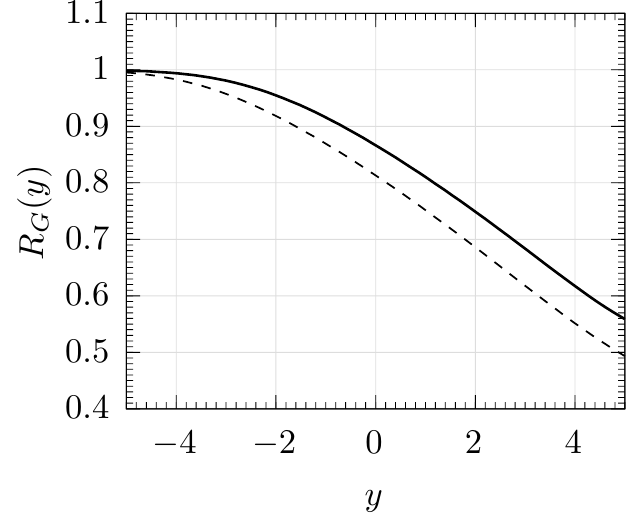}
    \caption{\label{fig:RG-b-y}
         Gluon shadowing factor $R_G(b_A)$ for photo-production of $J/\psi$ on lead as function of impact parameter $b_A$ (left), and rapidity $y$ (right). Solid and dashed curves correspond to c.m. collision
         energies $\sqrt{s_N}=5.02\TeV$ and $13\TeV$, respectively.}
\end{figure}

As an example, the magnitudes of $R_G$ at c.m. collision 
energies $\sqrt{s_N}=5.02\TeV$ and $13\TeV$
are presented in Fig.~\ref{fig:RG-b-y}  at $y=0$ 
as function of nuclear impact parameter $b_A$
in left panel, and vs rapidity for the $b_A$-integrated cross section in the right panel.
%

\subsection{Suppression of multi-gluon Fock states}
\label{n-g}

The lifetime of the $\bar QQg$ fluctuation, usually called coherence time (or length), in the nuclear rest frame is given by,
\BE
l^g_c=\frac{2E_{\gamma}}{M_{\bar QQg}^2}\,,
\label{lg}
\EE
where replacing $\la k_T^2\ra$ by $m_g^2$, the invariant mass of the fluctuation gets the form,
\BE
M_{\bar QQg}^2=
\frac{M_{\bar QQ}^2+k_T^2}{1-\alpha_g}+\frac{m_g^2+k_T^2}{\alpha_g}
\approx M_{\bar QQ}^2\left(1+\gamma/\alpha_g\right)\,.
\label{MQQG}
\EE
Here $\alpha_{g}$ 
is the fractional LF momentum carried by the gluon, which 
is small (see below), and 
\BE
\gamma=\frac{2m_g^2}{M_{\bar QQ}^2}\,.
\label{gamma}
\EE

Thus,  Eq.~(\ref{lg}) acquires the form,
\BE
l^g_c=\frac{P_g}{x\,m_N}\,,
\label{lg-new}
\EE
where according to \cite{Kopeliovich:2000ra} the quantity $1/x\,m_N$ is the maximal possible coherence length, and the factor 
\BE
P_g=\frac{\alpha_g}{\alpha_g+\gamma}\,,
\label{Pg}
\EE
considerably reduces the magnitude of $l_c^g$.
We simplify the calculations done in \cite{Kopeliovich:2000ra}, because we need to evaluate only the coherence lengths, including more complicated multi-gluon Fock components. We avoid the integration over 
the gluon-$\bar QQ$ transverse separation, fixing it at the mean value $1/m_g$.
Averaging $P_g$ over the radiation spectrum 
$d\alpha_g/\alpha_g$ we get its mean value,
\BE
\la P_g\ra 
=
\frac{\ln\left[(\alpha_g^{max}+\gamma)/(\alpha_g^{min}+\gamma)\right]}
{\ln(\alpha_g^{max}/\alpha_g^{min})}\,.
\label{mean-lg}
\EE

The minimal value $\alpha_g^{min}$ is defined by the trasnverse mass of the radiated gluon $\approx 2m_g$ and the available energy range for radiation. The latter is not well defined, it depends on the chosen invariant mass of the diffractive excitation $\gamma^*\to X$, corresponding to the triple-Pomeron term  in Gribov corrections.
Fortunately, the result of (\ref{mean-lg}) has a weak sensitivity (only Log) to this choice. Following 
\cite{Kopeliovich:2000ra} and previous studies of diffraction \cite{kklp} we fix
\BE
\alpha_g^{min}=\frac{2m_g^2}{0.2\,W^2}=
5\,\gamma\, x\,.
\label{al-min}
\EE
Here the chosen factor $0.2$ corresponds to the cut $x_F>0.8$ for Feynman $x$ in diffraction $hN\to hX$ (see e.g. in \cite{kklp}).

The choice of the upper bound $\alpha_g^{max}$ for the photon LF momentum is rather uncertain. Large values $\alpha_g^{max}\to1$ are dangerous, because according to Eq.~(\ref{MQQG}) the mass of the fluctuation rises to infinity. That means a vanishingly small probability of such fluctuations. Therefore, following \cite{Kopeliovich:2000ra} we fix $\alpha_g^{max}=\gamma$. Then, the numerator of (\ref{mean-lg}) equals to $\ln2$, and 
e.g. at $x=10^{-3}$ the mean value of the coherence length $\la l_g\ra$ is reduced by the factor $\la P_g\ra=0.13$, pretty close to the more accurate estimate in \cite{Kopeliovich:2000ra}.

Notice that at $\alpha_{min}\ll\alpha_{max}$ the invariant 
mass Eq.~(\ref{MQQG}) is dominated by the second term. 
Including more gluons to a Fock component (\ref{MQQG}) makes it heavier adding a new term 
$\gamma/\alpha_{gi}$.
Assuming the mean values of this terms equal each other, we get the 
coherence length for the $n$-gluon Fock component $n$ times shorter than the single-gluon 
$l^g_c$, which we already found very short.

This is why the Balitsky-Kovchegov (BK) equation \cite{balitsky,kovchegov}, which assumes that the dipole sizes are "frozen",  cannot be applied to nuclear targets at available energies \cite{poetic}.  Therefore 
the single-gluon approximation described in Sect.~\ref{1-g}, is sufficiently accurate, and we employ it in what follows. 

%
%
%
\section{Predictions and comparison with data}
\label{sec-data}
%
%
%

The dipole model calculations of diffractive photo-production of $J/\psi$ on protons, $d\sigma^{\gamma^\ast p\to \Jpsi p}(t)/dt$, were verified in Ref.~\cite{Kopeliovich:2021dgx} by comparing with data from the ZEUS \cite{Breitweg:1997rg,Chekanov:2002xi} and H1 \cite{Aktas:2005xu,Alexa:2013xxa} experiments at HERA. Predictions for charmonium photo-production on nuclei at high energies were provided in \cite{Ivanov:2002kc}.

Calculations, performed 
in the present paper, rely on the
quarkonium wave functions generated by the Buchm\"uller-Tye ({BT}) \cite{Buchmuller:1980su}
$Q$-$\bar Q$ interaction potential. Following \cite{Hufner:2000jb} we performed a Lorentz boost to the LF frame, including corrections for the effect of Melosh spin rotation. The Lorentz boost procedure was originally proposed in \cite{Terentev:1976jk} as an educated guess, but  proven later in \cite{Kopeliovich:2015qna} for a heavy $\bar QQ$ system.

Considering the LHC kinematic region, we included gluon shadowing calculated in the single gluon approximation, while multi-gluon Fock states were found to have too short coherence length and neglected. The corresponding suppression factor $R_g$ in Eq.~(\ref{eq:dipole:gs:replace-b}) provides  the main nuclear effect.   
The results of
calculations of the differential cross section for heavy vector meson production relying on the $\vec b$-dependent dipole models are presented and compared with available data in Fig.~\ref{Fig-tdep-data1}. 
\BF
\vspace*{-0.20cm}
\PSfig{1.10}{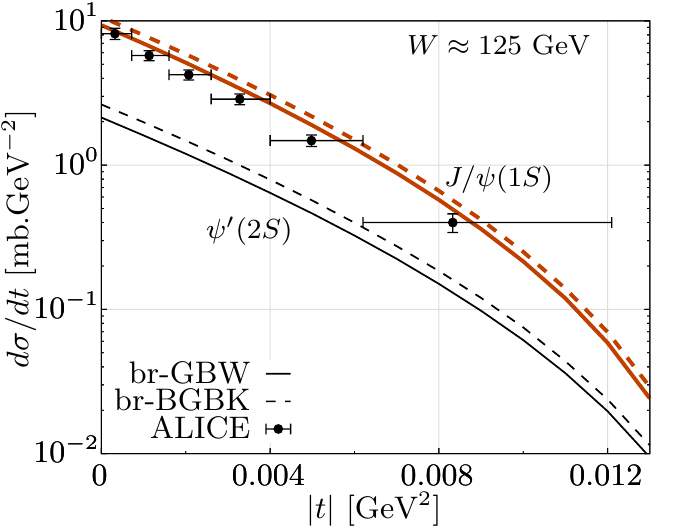}~~~
\PSfig{1.10}{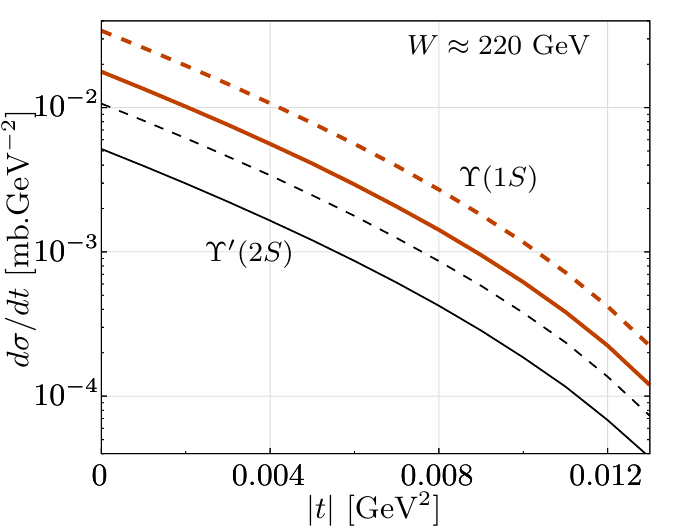}~~~~
\vspace*{-0.20cm}
\Caption{
  \label{Fig-tdep-data1}
(Color online) Predictions for the differential cross section of coherent photo-production of heavy quarkonia $d\sigma^{\gamma Pb\to V Pb}/dt$ in comparison with data from the experiment ALICE \cite{Acharya:2021bnz}. Predictions are depicted for the 1S (thick lines) and 2S (thin lines) charmonium (left panel) and bottomonium states (right panel). For the partial dipole-nucleus amplitude we employ the parametrizations br-GBW (solid lines) and br-BGBK (dashed lines).
}
\EF
The photo-production differential cross section $\gamma Pb\to \Jpsi Pb$ was extracted in \cite{Acharya:2021bnz} from UPC data at c.m. collision energy $\sqrt{s_N}= 5.02\TeV$ and at the mid rapidity, which corresponds to the c.m. $\gamma$-$N$ energy $W = \sqrt{M_V}\cdot(s_N)^{1/4}\approx 125\,\GeV$. They are depicted in the left panel of Fig.~\ref{Fig-tdep-data1} together with our predictions, which used two different $\vec b$-dependent dipole models. 
Our calculations are in a good accord with data.

In a wider interval of transverse momentum one should anticipate a specific diffractive behavior of the cross section with intermittent maxima and minima, as was predicted in \cite{Ivanov:2002kc}, and confirmed by the current calculations depicted in 
Fig.~\ref{Fig-tdep-data1}, which are based on more advanced models for the dipole amplitude. 
\BF
\vspace*{-0.2cm}
\PSfig{1.10}{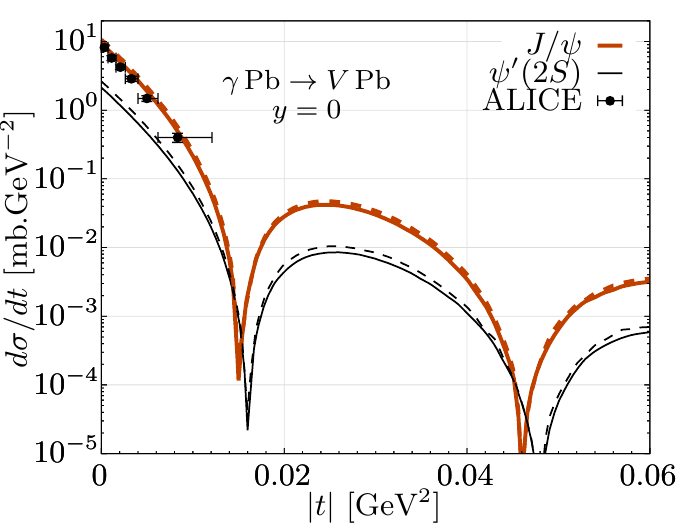}~~~~
\PSfig{1.10}{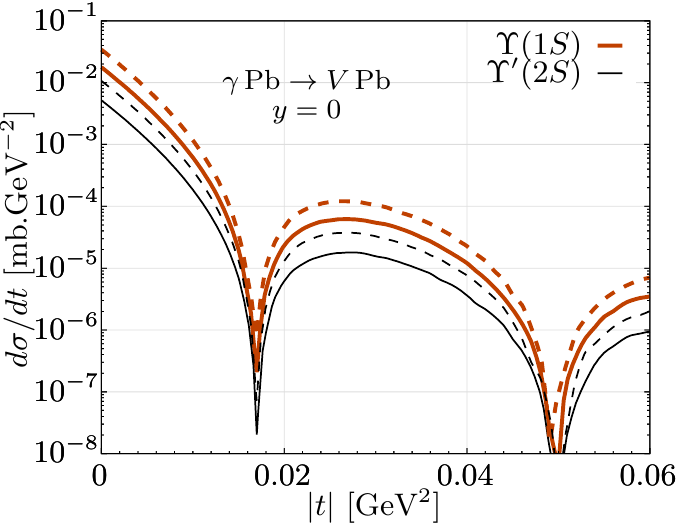}~~~~
\vspace*{-0.20cm}
\Caption{
  \label{Fig-tdep-data3}
 (Color online)  The same as in Fig.~\ref{Fig-tdep-data1}, but 
   for a wider $|t|$-range, at c.m. energy $W\approx \sqrt{M_V}(s_N)^{1/4}$ corresponding to the kinematic regions of UPC at the LHC at $y=0$. Our predictions are based on the br-GBW (solid lines) and br-BGBK (dashed lines)  parametrizations of the dipole cross section. 
  }
\EF
The curves show our predictions, still awaiting for data, for $d\sigma/dt$ of coherent photo-production of other quarkonium states, charmonium $\psip(2S)$ (left panel) and bottomonia $\Y(1S)$, $\Yp(2S)$ (right panel).  One can see that the production of the 1S and 2S bottomonium states is more sensitive to the choice of the model for partial dipole-proton amplitude in comparison with charmonium production. The differences in predictions adopting br-BGW and br-BGBK models can be treated as a measure of theoretical uncertainty.

%
%
%
\section{Summary}
\label{sec-sum}
%
%
%

We studied the momentum transfer dependence of differential cross sections for coherent photo-production of heavy quarkonia on nuclei, in the framework of the dipole description.

The LF wave function of a high-energy photon was expanded over Fock states, independently contributing to heavy quarkonium production, $|\bar QQ\ra$, $|\bar QQg\ra$, $|\bar QQ2g\ra$, etc.
The cross section $\gamma A\to VA$ was calculated for every Fock component separately in accordance with the corresponding coherence lengths. At the energies of UPC the photon energy in the nuclear rest frame is so high, that the coherence length, associated with the $\bar QQ$ component is much longer than the nuclear size, so one can eikonalize the dipole $\bar QQ$-N amplitude, as is done in Eq.~(\ref{eik}),
including the correlation between $\vec b$ and dipole orientation $\vec r$. Such a dipole amplitude contains Gribov corrections in all orders. The corresponding quark shadowing is a higher twist effect, so it is small at the scale imposed by the heavy quarkonium mass.

For the dipole amplitude $\gamma\to\bar QQ\to V$ we rely on  a  LF wave function of the  vector meson, determined by a solution of the Lorentz boosted Schr\"odinger equation with realistic potentials. On the other hand, we found that the frequently used unjustified model of photon-like structure $V\to\bar QQ$ leads to an exaggerated weight of the $D$-wave in the rest frame wave function, inconsistent with  the solutions of the Schr\"odinger equation.

The QCD dipole formalism also includes the leading twist gluon shadowing, which is related to the higher Fock components of the photon, $\bar QQg$, $\bar QQ2g$, etc. The corresponding shadowing effect is a leading twist due to large, nearly scale independent size of the $\bar QQ$-$g$ dipoles. The related nuclear effect is much stronger that the higher twist shadowing, controlled by the small-size of $\bar QQ$ dipoles. However, calculations of the effect of gluon shadowing is more involved, because the dipoles $\bar QQ-g, ...$ cannot be treated as "frozen" even at very high energies, their size fluctuates during propagation through the nucleus, and we took that into account by applying the path-integral technique.

The coherence length of the higher Fock components is much shorter than for the $\bar QQ$ fluctuation of the photon. We demonstrated that radiation of every additional gluon significantly reduces the coherence length. Therefore, at available energies the single gluon approximation was found to be rather accurate, while higher components containing two and more gluons have too short coherence length to produce a sizeable  shadowing effect. In particular, the BK equation cannot be applied to nuclear targets, because it treats all dipoles as "frozen".

On the contrary to the global data analyses, which provide only $b_A$ integrated gluon shadowing, we calculated the $b_A$ dependence of shadowing, which is crucial for the differential cross section $d\sigma^{\gamma A\to V A}/dt$.  

Our calculations of $d\sigma/dt$ for the coherent process $\gamma Pb\to\Jpsi Pb$ are in a good accord with recent ALICE data at the LHC (see Fig.~\ref{Fig-tdep-data1}).  We also provided  predictions for other quarkonium states, $\psip(2S)$, $\Y(1S)$ and $\Yp(2S)$ (see Figs.~\ref{Fig-tdep-data1} and \ref{Fig-tdep-data3}), that can be verified in the current experiments at the LHC.
%

\begin{acknowledgements}
This work was supported in part by ANID-Chile PIA/APOYO AFB180002.
The work of J.N. was partially supported by Grant
No. LTT18002 of the Ministry of Education, Youth and
Sports of the Czech Republic,
by the project of the
European Regional Development Fund No. CZ.02.1.01/0.0/0.0/16\_019/0000778
and by the Slovak Funding Agency, Grant No. 2/0020/22.
The work of M.K. was supported by the project of the International Mobility of Researchers - MSCA IF IV at CTU in Prague 
CZ.02.2.69/0.0/0.0/20\_079/0017983, Czech Republic.
\end{acknowledgements}



\end{document}